\documentclass[10pt,letterpaper]{article}

\usepackage{cogsci}
\usepackage{graphicx}
\usepackage{url}
\usepackage{subfig}
\usepackage{textcomp}
\usepackage{amsmath}

\newtheorem{assumption}{Assumption}

\cogscifinalcopy % Uncomment this line for the final submission 

\usepackage{pslatex}
\usepackage{apacite}
\usepackage{float} % Roger Levy added this and changed figure/table
\title{Reconciling Different Theories of Learning with an Agent-based Model of Procedural Learning}
 
\author{{\large \bf Sina Rismanchian (srismanc@uci.edu)} \\  
School of Education, University of California, Irvine \\
  Irvine, CA 92697 USA
  \AND {\large \bf Shayan Doroudi (doroudis@uci.edu)} \\
  School of Education, University of California, Irvine \\
  Irvine, CA 92697 USA}

\begin{document}

\maketitle

\begin{abstract}
  Computational models of human learning can play a significant role in enhancing our knowledge about nuances in theoretical and qualitative learning theories and frameworks. There are many existing frameworks in educational settings that have shown to be verified using empirical studies, but at times we find these theories make conflicting claims or recommendations for instruction. In this study, we propose a new computational model of human learning, Procedural ABICAP, that reconciles the ICAP, Knowledge-Learning-Instruction (KLI), and cognitive load theory (CLT) frameworks for learning procedural knowledge. ICAP assumes that constructive learning generally yields better learning outcomes, while theories such as KLI and CLT claim that this is not always true. We suppose that one reason for this may be that ICAP is primarily used for conceptual learning and is underspecified as a framework for thinking about procedural learning. We show how our computational model, both by design and through simulations, can be used to reconcile different results in the literature. More generally, we position our computational model as an executable theory of learning that can be used to simulate various educational settings.

\end{abstract}
\section{Introduction}
% Simulations of computational models of human learning can play a vital role in achieving different purposes such as teacher training, learning by teaching, and generating or evaluating hypotheses about human learning \cite{kaser2023simulated}. 
% To study learning from a computational point of view, in recent years, there has been a call from researchers to adopt a complex systems approach \cite{jacobson2006complex}. From this perspective, learning is best viewed as something that \textit{emerges}, rather than something that \textit{is}. 
% To understand learning as situated or social processes where interaction and collaborative work are prominent, complex systems can be insightful \cite{jacobson2016conceptualizing, abrahamson2007classroom}.
% One of the most prominent methods for exploring complex systems is agent-based modeling (ABMs) which is a bottom-up approach that is useful for understanding how micro-level behaviors give rise to macro-level emergent phenomena. 
% Although educational methodologies like ABMs of learning have been advocated, the literature features a limited number of such models, mostly illustrating specific learning scenarios \cite{abrahamson2005piaget,abrahamson2007classroom}. In this work, we implement an ABM that reconciles ICAP with two prominent cognitive frameworks of learning.

The ICAP framework \cite{chi2014icap} is one of the most influential recent theories of active learning as it is the most cited paper in \textit{Educational Psychologist} since the date it was published. In this framework, Chi and Wylie suggest that there are four different levels of cognitive engagement (\textit{passive}, \textit{active}, \textit{constructive}, and \textit{interactive}) and that higher modes of engagement result in greater learning outcomes. While research has shown that the ICAP hypothesis holds in a variety of studies, there are several limitations to the ICAP framework.

First, although it is not made explicit, ICAP primarily makes sense in the context of conceptual learning, especially when it comes to constructive and interactive learning. The authors {\it do} describe how self-explanation could be a constructive activity in the context of procedural learning; but even then, what is learned from self-explanation is ``rationales and justifications,'' a form of conceptual understanding \cite{chi2014icap}.

Second, related to the first point, there are many kinds of learning where we expect the ICAP hypothesis to not apply. This was admitted in the original ICAP paper: ``for domains or topics in which students cannot bootstrap themselves into deeper understanding due to a lack of relevant schemas, or for topics for which no deeper rationales exist, the predictions made by ICAP may not hold'' \cite{chi2014icap}. In particular, the authors describe how learning English articles (with somewhat arbitrary rules) could be one such domain. More generally, the Knowledge-Learning-Instruction (KLI) framework provides a formal taxonomy of knowledge components (KCs), only some of which are best learned using ``understanding and sense-making processes'' \cite{klikoed2012}. Therefore, it seems as though ICAP cannot account for many kinds of realistic learning scenarios. While KLI might give guidelines for what these scenarios are, these two frameworks have not been juxtaposed in an easy-to-interpret way to our knowledge.

Finally, some theoretical frameworks, such as cognitive load theory (CLT) \cite{sweller2011cognitive}, actually make predictions that oppose the ICAP hypothesis. As mentioned by the authors \cite{chi2014icap}: \begin{quote}
    ICAP’s predictions seem to be the opposite of predictions from the [cognitive] load theory. For example, cognitive load theory states that the greater the load in the to-be-processed presented materials, the more difficult a task becomes, and therefore the less resulting learning. However, ICAP makes the opposite prediction. 
\end{quote} 
CLT has primarily been studied in the context of procedural learning tasks, which is possibly why the two theories make opposing predictions in some cases, but each theory has decent explanatory power in the areas they are primarily applied in (conceptual learning for ICAP and procedural learning for CLT).

To help address all of these issues, we devised a novel computational model of learning: Procedural ABICAP. Computational models can play a vital role in achieving different purposes such as teacher training, learning by teaching, and generating or evaluating hypotheses about human learning \cite{kaser2023simulated}. Specifically, we developed an agent-based model of procedural learning drawing upon the ICAP framework. Our work builds upon the ABICAP model, an agent-based model rooted in ICAP \cite{rismanchian2023computational}. However, given the focus of ICAP, the ABICAP model was also primarily designed to account for conceptual learning. Moreover, this model did not account for CLT. Therefore, we sought to see if we could develop a version of ABICAP that could account for (1) learning procedural knowledge, (2) the predictions made by KLI, and (3) the predictions made by CLT.

Constructing a computational model forces us to be more explicit and precise about the predictions of a model {\it in different learning situations}. In this paper, we focus on the theoretical ability of the model to unify different theoretical frameworks of learning. However, the true potential of the model, which we briefly describe at the end, is that it could serve as an {\it executable} theory that could generate new hypotheses about learning in different educational settings.

We begin by describing the theoretical frameworks that we draw upon: namely ICAP, KLI, and CLT. Then we describe the model, including the key theoretical assumptions that we make in this work. We then describe several simulations using Procedural ABICAP to show how it could help reconcile the different theoretical frameworks and replicate existing prior empirical results. We end with a discussion of the potential utility of such a model in producing new insights into the nature of teaching and learning.

\section{Related Work}
\subsection{ICAP}
According to ICAP, in the passive mode, students passively store the information that they are being taught about and thus gain minimal understanding. For instance, a passive activity could be listening to a lecture without doing anything else. Active learning is a higher mode of engagement in which the new information activates students' prior knowledge and the student integrates the activated prior knowledge with new knowledge. For instance, when the learner takes notes during a lecture, they are learning actively. The constructive mode is a higher mode of engagement in which the learner constructs new knowledge that goes beyond the information given by integrating the new information with activated prior knowledge. For instance, a student who draws a concept map gains a deeper understanding which can potentially transfer to related topics they had not previously learned. Finally, interactive learning takes place when a group of two or more learners who are already learning constructively build on each others' knowledge through interaction. For instance, this could be when collaboratively drawing a concept map in a dyad. Based on a large body of empirical literature, ICAP hypothesizes that the following relationship holds in terms of learning outcomes:
$$\text{Interactive} > \text{Constructive} > \text{Active} > \text{Passive}$$

% Although ICAP is a prominent framework, there are some limitations in the framework that need to be addressed. To the best of our knowledge, ICAP is mostly studied for learning conceptual knowledge rather than procedural knowledge. In fact, as we discuss below the ICAP framework seems to mostly make sense for learning conceptual knowledge, rather than learning procedures or even simple declarative knowledge. 

% For example, Menekse et al. \cite{menekse2013differentiated} examine ICAP's hypotheses in the domain of metallurgy where the knowledge learned by students is basically conceptual. This paper serves as pivotal empirical evidence for the ICAP framework. 

\subsection{KLI}
The KLI framework articulated by Koedinger et al. \cite{klikoed2012} tries to determine instructional practices that best serve specific knowledge components and learning events. The framework tries to address opposing recommendations based on different frameworks of learning; for example, while some researchers advocate for ``instruction that increases demands on students so as to produce desirable difficulties,'' other researchers draw on cognitive load theory to advocate for decreasing the extraneous cognitive load on students \cite{sweller2011cognitive}. KLI tries to address these oppositions by distinguishing between different types of knowledge components (KCs) and learning events to recommend optimal instructional practices for each setting. 

Different types of KCs are distinguished in KLI by several variables: the kind of input (constant or variable), the kind of response (constant or variable), whether the relationship between the input and output can be verbalized, and whether it has some kind of rationale. Moreover, there are three kinds of learning processes in KLI: memory and fluency-building, induction and refinement, and understanding and sense-making. The first two are generally useful for learning procedural KCs while the third one is useful for learning conceptual KCs. For instance, learning Chinese vocabulary involves learning a set of largely isolated facts that (typically) do not have a rationale\footnote{While previous work has categorized Chinese vocabularies as KCs that are not rationale, it is worth mentioning that it is not always the case, as some types of rationales could be derived from pictographs, etc. For the purpose of this paper, we use these KCs as an example, while other types of KCs can be of similar types, such as learning the capitals of countries (i.e., France $\rightarrow$ Paris).}; ``while a few radicals may have iconic value, the relationship generally has no rationale, instead being a writing convention'' \cite{klikoed2012}. These kinds of KCs are best supported by memory and fluency-building activities. On the other hand, KCs that have a rationale (e.g., steps in a math problem or aspects of a scientific concept) are supported by understanding and sense-making processes. We believe the latter are the kinds of KCs that ICAP typically focuses on.
% Interestingly, the works on the role of instructional activities such as prompted self-explanation which has largely been conducted by Chi and colleagues \cite{CHI1994439} have shown to be optimal instruction for learning processes that are categorized under understanding and sense-making. 

% and fluency as it comes from a constant -> constant knowledge component, whereas using English articles comes from a variable -> constant knowledge component is categorized under induction and refinement, and finally, learning how to use plural English spellings comes from a variable -> variable KC which leads it to be categorized under understanding and sense-making. 

\subsection{Cognitive Load Theory}
Cognitive load theory suggests that learning is most effective when cognitive processing is kept to a manageable level, as human working memory has limited capacity \cite{sweller2011cognitive}. It posits that instructional designs should minimize extraneous cognitive load to facilitate the encoding of information into long-term memory. Two main implications of CLT are the worked example effect and the expertise reversal effect. The worked example effect states that learners with low prior knowledge can benefit more from studying worked examples rather than solving problems themselves, as it reduces extraneous cognitive load, allowing them to focus on understanding the underlying principles and processes \cite{sweller2006worked}. The expertise reversal effect expands upon this by noting that ``Instructional techniques that are highly effective with inexperienced learners can lose their effectiveness and even have negative consequences when used with more experienced learners'' \cite{kalyuga2003expertise}.
For example, in the domain of algebra equations, it has been shown that studying worked examples is more beneficial than problem-solving for novice learners (i.e., worked example effect) whereas, for learners with sufficient prior knowledge, problem-solving is associated with better outcomes compared to studying worked examples \cite{sweller2006worked}. These findings are inconsistent with ICAP; based on ICAP, we would anticipate that students who solely engage in problem-solving (active or constructive) have better learning outcomes than the ones who study worked examples (passive). While, Chi and Wylie do not directly address the worked example effect, they suggest that some of the benefits of worked examples may be due to explicit self-explanation (a constructive activity) or implicit self-explanation when having students complete some of the steps in an incomplete worked example. They also note that fading worked examples (gradually requiring students to complete more steps) has been shown to be more effective as students develop more proficiency. Clearly, this last observation suggests that the benefit of constructive activities is somewhat dependent on expertise, but they do not give a theoretical explanation for why this is from within the ICAP framework.

\section{Procedural ABICAP}
Based on the existing differences in the recommendations of these frameworks, we develop an agent-based model of procedural learning. Our model, Procedural ABICAP, can serve as an executable theory of learning that is designed based on existing theoretical frameworks. By concretely instantiating theoretical assumptions in a computational model, we can potentially clarify (or problematize) certain aspects of the underlying theories; in this case, we hope to resolve some of the differing claims made by different frameworks. Moreover, a practical contribution of such executable models is that they are able to simulate specific settings of learning for a variety of domains and can thus be used to generate new hypotheses that could be verified using further empirical studies.

The model is largely based on the recently-proposed open-source ABICAP model \cite{rismanchian2023computational} with a few changes made to the model in order to (1) make it reasonable for procedural learning and (2) help reconcile across ICAP, CLT, and KLI. The first change is that the ABICAP model assumed that knowledge of each piece of knowledge is binary: a learner either learns something or does not. This may be reasonable for conceptual knowledge (e.g., where each node is a small piece of declarative knowledge connected to other declarative knowledge); however, for procedural knowledge, it seems more realistic for learning to take place gradually over time as evidenced by learning curves \cite{klikoed2012}. Moreover, this seems to be a corollary of the expertise-reversal effect: if the optimal learning activity for a particular skill depends on a learner's degree of expertise on that skill, then that suggests mastery gradually develops over several practice opportunities.

The other two ways in which we modified the model form two major assumptions underlying Procedural ABICAP. We believe these assumptions (whether right or wrong) can enhance theoretical discourse in the cognitive sciences in service of reconciling different theories:

\begin{assumption}[Cognitive Load]
Different modes of engagement have different levels of cognitive load. In particular cognitive load increases as the degree of engagement increases (i.e., $cl_{\text{passive}} < cl_{\text{active}}  < cl_{\text{constructive}} < cl_{\text{interactive}}$).
\end{assumption}

\begin{assumption}[Procedural Knowledge Construction]
When a learner practices a procedural skill constructively or interactively, they do not simultaneously learn other skills, but rather they can increase in their ability to connect the practiced skill to related skills. That is, knowledge does not immediately transfer to other related skills, but rather constructive learning acts as a form of ``preparation for future learning'' \cite{bransford1999chapter} of related skills.
\end{assumption}

We expect the first assumption to be relatively uncontroversial, as it is almost implied by CLT. However, as we describe below, this assumption can act as a way to reconcile results in ICAP and CLT.

The second assumption is more speculative and debatable. Our rationale behind this assumption is that constructive learning is most natural for conceptual learning; that is students construct their {\it understanding} of a domain. At first glance, this does not make sense in the context of procedural knowledge. One way to reconcile this is that learners construct conceptual knowledge in relation to the procedural skill they are practicing. Indeed, this was hinted at by Chi and Wylie: ``For procedural domains, self-explaining has been traditionally implemented in a constructive way in the context of a worked example by having students explain each example step to themselves, such as justifying how a solution step follows from a prior step.'' In essence, we think this kind of conceptual knowledge (e.g., ``justifying how a solution step follows from a prior step'') is {\it relational}, connecting two procedural skills. In other words, in our model, conceptual knowledge is implicitly modeled as the connection between procedural knowledge. We believe this is a novel way of theorizing conceptual and procedural knowledge representation. However, we also believe this could be an oversimplification, and we describe what we believe may be a better way of integrating conceptual and procedural knowledge at the end of this paper.

We note that the particulars of our model as presented below are certainly open to debate. By presenting a concrete model, we hope to initiate productive discussions on the specific mechanisms underlying human learning. We especially note that the choice of constants is arbitrary and may need to be calibrated for particular settings. Specific constants were chosen to ensure learning operates on a reasonable time scale and to depict clear contrasts between different modes of learning on those time scales. However, for the particular analyses we focus on in this paper, changing the constants does not significantly change the qualitative nature of the results.

\subsection{Knowledge}
Procedural ABICAP assumes that there is a set of knowledge components modeled as nodes in a knowledge graph. Each node in the knowledge graph represents a procedural knowledge component that a student has mastery over. Also, knowledge components can be connected to each other via undirected edges. For instance, the edge that connects two nodes $i$ and $j$ is named $w_{ij}$. Each edge has a default associated weight which shows the extent of the relationship between two nodes. As per Assumption 2, edge weights may be increased in the learning process for constructive and interactive learners but will remain unchanged in the learning of active and passive learners.

For example, for solving algebra equations, each node represents a typical step that a student may take in solving the problem. As an example, if a student tries to solve $2x + 5 = 6x - 7$, then there are multiple nodes such as adding $2x$ to both sides of the equation to isolate the $x$ terms on one side (resulting in $5 = 4x - 7$),  adding $7$ to both sides (resulting in $12 = 4x$), and dividing both sides by $4$ to finding the value of $x$. In this example, the steps that are taken consecutively might have higher edge weights compared to steps that are not consecutive. Moreover, steps may have edges to steps not involved in this problem (e.g., dividing both sides of the equation might have connections to division and fractions problems outside of equations).

The knowledge graph structure varies based on the domain of the knowledge. For instance, for learning KCs that have no rationale, the knowledge graph may simply consist of a large collection of nodes (e.g., Chinese vocabulary words) with no or few edges connecting them. On the other hand, for solving algebra equations, students can learn the relationships between the steps and the meaning of each one, so the knowledge graph may have many edges between steps that appear in related problems and steps that have prerequisite relationships. 

\subsection{Learning}
We model learning for each mode of engagement based on the proposed knowledge change process of each mode as described in the ICAP framework \cite{chi2014icap}. In each step of our model, learners \textit{practice} one node named the practicing node. Practicing a specific node in the model is  described by the following function:

\begin{equation}
\label{eq:learning}
M_{t + 1, i} = \sigma\Biggl(3.5 \times \Bigl(M_{t, i} + \delta \sum_{j} M_{t, j}\, w_{ij,t}\Bigr)\\
- \bigl(b_i + cl_{\text{mode}}\bigr)\Biggr)
\end{equation}

where 
\begin{align*}
    0 \leq w_{ij,t} \leq 1 , \qquad \forall t, \forall i \in [1, N_{KCs}], \quad 0 \leq M_{t, i} \leq 1
\end{align*}
where $M_{t, i}$ represents the mastery over node $i$ at time step $t$, $\sigma$ is the sigmoid function, $\delta$ is 0 if the mode is passive and 1 otherwise, $w_{ij,t}$ is the edge weight for the edge between nodes $i$ and $j$ at time $t$, $b_{i}$ represents the difficulty of learning node $i$ (which could be due to a variety of factors, including {\it item-specific} intrinsic cognitive load), $N_{KC}$ is the number of nodes or knowledge components, and $cl_{\text{mode}}$ represents the cognitive load for the current mode of engagement. We set $b_i$ to be the same for all nodes in this study, as we are not attempting to model item-specific differences and we do not expect varying item difficulties to affect our overall results. We also make $cl_{\text{mode}}$ increase monotonically as the mode of engagement increases as per Assumption 1. The sigmoid function exhibits a characteristic ``S'' shape, where it starts with a sharp increase, transitioning smoothly to a plateau as the input value increases, reflecting its nature as a squashing function that limits its output to values between 0 and 1. 

% for learners who are not passive (i.e., $\text{mode} \in \{\text{active, constructive, interactive}\}$). 

% If the learner is passive, then the function is:
% \begin{equation}
% \label{eq:passive}
%     \mastery_{t + 1, i} = \sigmoid(3.5 \times \mastery_{t, i} - (b_i + cl_{\text{passive}}))
% \end{equation} 

Although Equation~\ref{eq:learning} underlies how a learner's mastery over the practicing node increases under each of the four modes, there are mode-specific aspects of the learning process which further differentiate the different modes of engagement, as we describe below.

\begin{figure}[t]
    \centering
    \includegraphics[width=8cm]{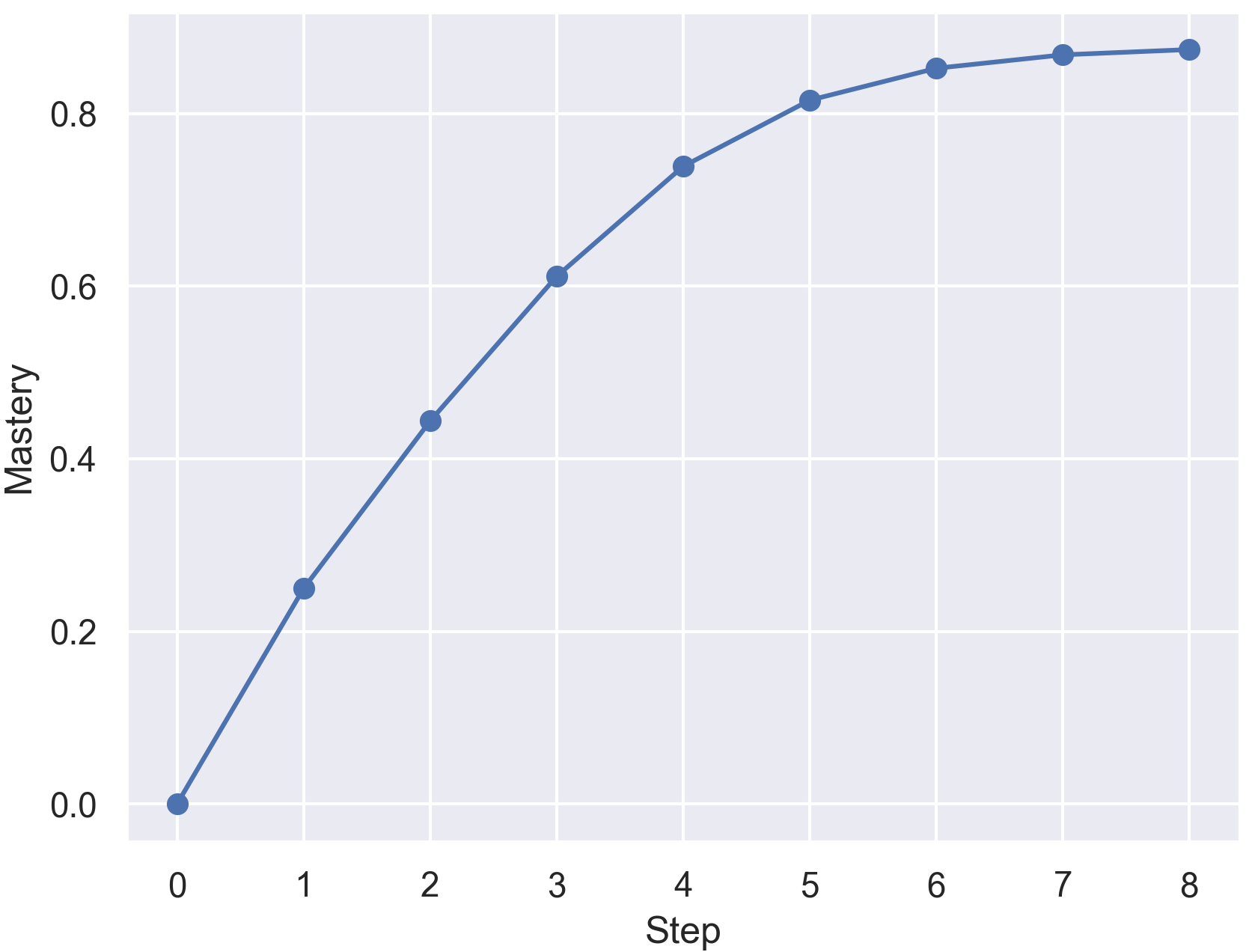}%
    \caption{Passive learner's mastery over a node through steps}
    \label{fig:passive}%
\end{figure}

\paragraph{Passive}
\textit{Passive} learning is described as a process of storing knowledge in an {\it isolated} way \cite{chi2014icap}. In procedural learning, this translates to gaining mastery over a node regardless of other nodes. Conversely, gaining mastery in passive nodes should not affect the other nodes. As shown in equation \ref{eq:learning}, the new mastery is only a function of the previous mastery over the same node (since $\delta_{\text{passive}} = 0$), the cognitive load of passive learning, and the difficulty of the node. Passive learning has the lowest cognitive load \cite{chi2014icap,klikoed2012} compared to other modes of engagement. Learning of a typical passive learner is shown in Figure~\ref{fig:passive}, where with five practice opportunities, the learner achieves mastery over a node.

% \begin{figure}[t]
%     \centering
%     \includegraphics[width=7cm]{figs/sigmoid_plot.png}
%     \caption{Passive learner's mastery over a node through steps}%
%     % \label{fig:passive}%
% \end{figure}

\begin{figure}[t]
    \includegraphics[width=8cm]{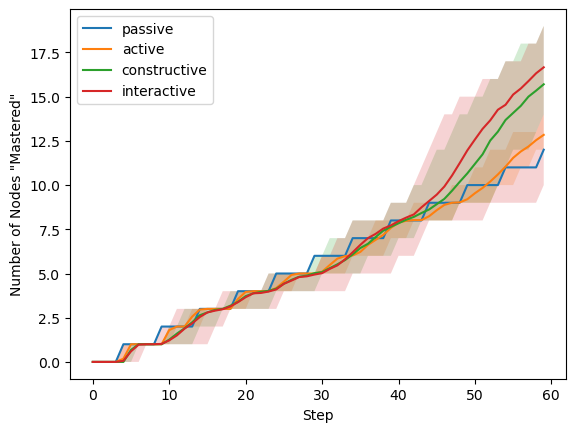}
    \caption{Simulation with 20 nodes that demonstrates the expertise reversal effect.}
    \label{fig:simulation}
\end{figure}

\paragraph{Active}
% Active learning is described as ``integrating'' processes where prior knowledge is activated and ``new information can be assimilated into the activated schema'' \cite{chi2014icap}.
As shown in Equation~\ref{eq:learning}, the impact of prior knowledge in \textit{active} learning is modeled by having the learning rate increase as a function of the mastery over each adjacent node multiplied by the weight of the edge that connects those nodes (since $\delta_{\text{active}} = 1$)

\subsubsection{Constructive}
% Constructive learning is described as learning that invokes processes such as comparing, contrasting, analogizing, generalizing, or self-explaining that not only requires integrating prior knowledge with new information but also involves ``inferring`` new knowledge that goes beyond the material presented to the learner. Although the description is best interpreted for conceptual knowledge, in the domain of procedural knowledge these kinds of processes could arise when, for instance, a learner infers the relationship between different steps of solving an algebra equation. 
In addition to the learning described by Equation~\ref{eq:learning}, we model \textit{constructive} learning in Procedural ABICAP by ``reinforcing'' edges between the practicing node and neighboring nodes, as implied by Assumption 2. Concretely, at each step of practicing, a constructive learner increases the edge weights of at most two arbitrary edges between the practicing node and neighboring nodes by a constant amount. When practicing node $i$, for a randomly chosen neighboring node $j$,
\begin{equation}
\label{eq:constructive}
w_{ij,t+1} = \min(w_{ij,t} + \Delta_{j,t+1}, 1)
\end{equation}
where $\Delta_{j,t+1} \sim \textsc{uniform}(0, 0.15)$. The same procedure is followed for another randomly chosen node that neighbors $i$. Note that if there are no edges between the practicing node and the other nodes, then the constructive learner cannot reinforce any edges.

\subsubsection{Interactive}
% Interactive learners learn by ``co-inferring'' processes that involve both partners participating in constructive learning (as described above) and taking turns sharing the knowledge that they learned and constructed with each other. As Chi and Wylie describe it ``This mutuality further benefits from opportunities [and] processes to incorporate feedback, to entertain new ideas, alternative perspectives, new directions, etc.'' \cite{chi2014icap}
We model practicing in the \textit{interactive} mode as follows. At the beginning, interactive learners are randomly paired into dyads. At each step, the partners practice node $i$ and reinforce edges as in constructive learning, except that the upper bound in how much they add to each edge weight is higher and dependent on their partner's edge weights. For partners $A$ and $B$ (where superscripts indicate edge weights for the associated partner),
\begin{eqnarray}
\label{eq:interactive}
w^A_{ij,t+1} = \min(w^A_{ij,t} + \Delta^A_{j,t+1}, 1) \\
w^B_{ij,t+1} = \min(w^B_{ij,t} + \Delta^B_{j,t+1}, 1)
\end{eqnarray}
where $\Delta^A_{j,t+1}, \Delta^B_{j,t+1} \sim \textsc{uniform}(0, 0.05 + \max(|w^A_{ij,t} - w^B_{ij,t}|,  0.15))$. The same procedure is followed for another randomly chosen node that neighbors $i$. In effect, if $A$ already has an edge weight that is more than 0.15 larger than $B$'s analogous edge weight, then $A$'s edge weight can increase by a larger amount (i.e., $A$ benefits from $B$'s greater knowledge about the conceptual knowledge associated with $i$).

\section{Simulations and Results}

In this section, we show how Procedural ABICAP can be used to account for different theoretical phenomena.

\subsection{Simple Procedural KCs}
Chi and Wylie admit that ``one type of topic for which increased cognitive engagement may not be helpful is simple procedural domains for which the rules are arbitrary and cannot be logically deduced.'' In KLI terms, this would be any KC that is non-verbal and where there is no associated rationale, such as Chinese vocabulary learning and learning when to use which English articles (an example mentioned in both the ICAP \cite{chi2014icap} and KLI \cite{klikoed2012} papers). According to KLI, this includes associations, categories, and productions (or arbitrary rules). Indeed, Wylie et al. demonstrated that when learning English articles, self-explanations do not appear to be helpful. Chi and Wylie suggest this may be because there is no rationale; Wylie et al. also explained this result ``could be that the act of generating and selecting explanations added extraneous cognitive load to the task.'' 

By design, Procedural ABICAP makes this prediction. If there is a knowledge graph with no edges, the benefit of learning or reinforcing edges in active, constructive, and interactive learning disappears. Moreover, passive learning becomes the best form of learning because the mastery equation for all four modes becomes equivalent (i.e., the summation in equation 1 disappears) except for $cl$ where passive learning has the smallest amount of cognitive load.

\subsection{ICAP Hypothesis and Expertise Reversal Effect}

In order to assess our model's ability to replicate the ICAP hypothesis and the expertise reversal effect, we rely on a simple simulation as follows. For the knowledge graph, we used an undirected Watts-Strogatz graph \cite{watts1998collective} with 20 nodes ($N_{KCs} = 20$), where each node has three edges on average. For each mode of engagement or experimental condition, we simulate 50 learners. Each learner practices the same node $i$ until they have $\text{Mastery}_{i} > 0.8$, at which point they move on to the next node. In our results, we report the average number of nodes ``mastered'' by agents in each condition (i.e., the average number of KCs $i$ where $M_{i} > 0.8$ for each group. Each simulation is run for 60 steps. The setting of variables is as follows: $ b_{i} = 0.6, cl_{\text{passive}} = 0.5, cl_{\text{active}} = 0.65, cl_{\text{constructive}} = 0.75, cl_{\text{interactive}} = 0.8$.

As shown in Figure~\ref{fig:simulation}, after enough steps, the ICAP hypothesis does indeed hold (i.e., $I > C > A > P$). However, what is more interesting is that this is not always the case. When learners have low prior knowledge, the ICAP hypothesis does not hold. In particular, until around step 35, we find that passive learning is consistently at least as good as (and typically better than) active and constructive learning. This is consistent with the worked example effect. However, after step 35, active and constructive learning gradually overtake passive learning, consistent with the expertise reversal effect. The reason for this is that early on, the increased cognitive load is too high to make active or constructive learning worthwhile, but as the learners learn more, the support of neighboring nodes (and the reinforcement of edges in the case of constructive learning) makes these kinds of learning activities more beneficial, {\it despite} the increased cognitive load. This shows why it could be important to take both cognitive load considerations and ICAP considerations into account to make accurate predictions about learning.

\section{Discussion and Future Work}
Our model represents a more nuanced version of ICAP that is designed to account for empirical results on learning procedural knowledge. Our results show that by modeling cognitive load, the ICAP framework may be able to explain prior empirical literature beyond results on purely conceptual learning. Here we discuss three directions we could pursue in future work to truly capitalize on the benefits of Procedural ABICAP.

\subsection{Replication Studies}

While our model can replicate some broad findings in the literature (e.g., the ICAP hypothesis and the expertise reversal effect), we note that these replication studies are still preliminary. In future work, we could do more precise replications of specific findings. For example, here we showed how conceptually our model could replicate the expertise reversal effect. However, we do not know if it would actually predict expertise reversal effects noticed in the literature. By trying to replicate specific studies that noticed such an effect (e.g., by mimicking their knowledge graphs and experimental setups), we may get better insights into the validity of our model and how it may need to be adjusted. Moreover, Procedural ABICAP could be used to replicate a variety of other results in the literature as well. For example, we could try to replicate results on studies that interleave worked examples and problem-solving tasks \cite{van2011effects}.

\subsection{Making Novel Predictions}

Although we have primarily focused on replicating prior results, the true power of agent-based modeling lies in making predictions about novel settings where there is little empirical literature. This could be used to generate hypotheses which could subsequently be verified with experimental studies \cite{rismanchian2023computational}. 

Researchers who design interventions based on the ICAP framework often assume that learning gains from higher modes of engagement are higher \cite{dimitrova2022choice} but as we have shown with our model, this is not always the case. Using a computational model could help elucidate when we expect certain instructional interventions to be more effective than others. Of course, for the simulations to be accurate, it would help to have a concrete sense of the knowledge graph, the cognitive load for different modes of engagement, etc. If researchers have a lot of uncertainty about certain parameters (e.g., the relative cognitive load for different activities), they can use robustness tests to assess the sensitivity of the simulation results to the choice of those parameters. Moreover, models like Procedural ABICAP could also support researchers who are trying to develop adaptive algorithms for sequencing different kinds of instructional activities (e.g., using reinforcement learning). For instance, researchers designing intelligent tutoring systems have used ICAP as a general framework to devise adaptive scaffolding for cognitive engagement \cite{fahid2021adaptively}. The authors found that, according to their models, adaptively choosing scaffolding activities is better than always choosing constructive scaffolds (i.e., the highest mode they could support based on ICAP). Our model can serve as a complementary theoretically grounded environment for comparing different data-driven approaches for adaptively sequencing activities or to gain insights into when algorithms should use different kinds of activities.

% Such models could also be useful for studying longitudinal studies or seeing the long-term effects of some interventions, which could be prohibitively expensive and difficult to do in practice \cite{abicap}.

\subsection{Bridging Between Conceptual and Procedural Learning}

While we claimed ICAP has primarily been studied in the context of conceptual learning, our model swings the pendulum to the opposite extreme of only focusing on procedural knowledge acquisition. We are currently working on developing a model that can account for both conceptual and procedural learning. This would be important for several reasons. First, we noted earlier that Assumption 2 (i.e., that knowledge construction is in terms of reinforcing edges in the knowledge graph) may be an oversimplification. In reality, we believe that constructively learning procedural skills leads to constructing conceptual knowledge. In Procedural ABICAP, conceptual knowledge {\it is} implicitly the edges between nodes. However, this is likely not sufficiently general. There may be a concept that connects more than two nodes, in which case constructing conceptual knowledge would transfer to several different procedural skills. The most natural way of incorporating this is by having both conceptual and procedural nodes in the knowledge graph. Second, it has been shown that learning procedural knowledge has a positive impact on learning corresponding conceptual knowledge and vice versa \cite{rittle2015not}. We could draw upon this body of literature to extend our model, which could in turn generate new insights into the relatively understudied bidirectional relationship between procedural and conceptual knowledge \cite{rittle2015not}.

\bibliographystyle{apacite}

\setlength{\bibleftmargin}{.125in}
\setlength{\bibindent}{-\bibleftmargin}

\bibliography{CogSci_Template}

\end{document}